\documentclass[aps,pre,preprint,
superscriptaddress,longbibliography]{revtex4-2}

\usepackage{amssymb}
\usepackage{graphics}
\usepackage{amsmath}
\usepackage{graphicx}
\usepackage{amsfonts}
\usepackage{newunicodechar}
\usepackage{xcolor} 
\usepackage{listings}
\usepackage{import}
\usepackage{lineno}
\usepackage{pgfplots}

\usepackage{multirow}
\usepackage{lmodern}
\usepackage{appendix}
\usepackage{longtable}

\usepackage[caption=false]{subfig}

\pgfplotsset{compat=1.18}

\newcommand{\SymPhas}{\textsc{SymPhas}}
\newcommand{\SPnew}{\textsc{SymPhas\,2.0}}

\newcounter{bla}

\usepackage{setspace}

\begin{document}

\title{\SPnew{}: Parallel and GPU accelerated code for phase-field and reaction-diffusion simulations}

\author{Steven A. Silber}
\affiliation{Department of Physics and Astronomy,  Western University, 1151 Richmond Street, London, Ontario, Canada  N6A\,3K7}

\author{Mikko Karttunen}
\email[]{mikko.karttunen1@uef.fi}
\affiliation{Department of Physics and Astronomy,  Western University, 1151 Richmond Street, London, Ontario, Canada  N6A\,3K7}
\affiliation{ELLIS Institute Finland, Maarintie 8, 02150 Espoo, Finland}
\affiliation{Department of Technical Physics, University of Eastern Finland, P.O. Box 1627, FI-70211 Kuopio, Finland}
\affiliation{Department of Chemistry,  Western University, 1151 Richmond Street, London, Ontario, Canada N6A\,5B7}

\date{\today}

\begin{abstract}
\singlespacing

We present \SPnew, a major update of the compile-time symbolic algebra simulation framework \textsc{SymPhas} for phase-field and reaction-diffusion models. This release introduces significant expansions and enhancements that enable the definition of a phase-field model directly from the free-energy functional via compile-time evaluated functional differentiation. It also  introduces directional derivatives, symbolic summation, tensor-valued expressions, and compile-time derived finite difference stencils of arbitrary order and accuracy. 
Furthermore, the code has been parallelized for CPUs with MPI, and
GPU computing has been added using CUDA (Compute Unified Device Architecture).
For the latter, symbolic expressions are compiled into optimized CUDA kernels, allowing large-scale simulations to execute entirely on the GPU. For large systems ($32,768^2$ in 2D and $1,024^3$ in 3D with double precision), speedups up to $\sim \!\!1,000 \times$ were obtained compared to the first version of \SymPhas{} using multi-threaded CPU execution on a single system. 
These developments establish \SPnew  ~as a flexible and scalable framework for efficient implementation of phase-field and reaction-diffusion models on GPU-based high-performance computing platforms.

\end{abstract}

\maketitle

\singlespacing

\section{Introduction}

Phase-field modeling is widely used in materials science for simulating microstructural evolution under a broad range of thermodynamic and kinetic conditions~\cite{Moelans2008}. The method represents the  phase behavior of a system using continuous order parameters that vary smoothly within bulk regions and rapidly across interfaces, thereby eliminating the need for explicit interface tracking~\cite{Hohenberg1977,Provatas2010} and providing both a qualitative and a quantitative description of microstructure growth~\cite{Chen2002}. This diffuse interface approach enables the simulation of diverse complex phenomena such as grain growth, solidification, fracture~\cite{Spatschek_2011}, biological membranes \cite{Fan_2008}, cellular systems \cite{Nonomura_2012,Palmieri2015}, and even immune response \cite{Najem2014}. Phase-field models are typically governed by systems of partial differential equations derived from a free energy functional.

Analogous approaches can be used to model reaction-diffusion systems, which are typically defined directly from their dynamical equations instead of a free energy functional. Reaction-diffusion systems are often used to describe spatial pattern formation in chemical and biological systems and media~\cite{Soh2010-rz,Landge2020-bf,Volpert2025-vi}. Some well-known models are the Gray–Scott~\cite{Gray_1985} and Turing models~\cite{Turing1952,Lengyel1992-di,Leppaenen2002, Kondo2010,Maini2012}, which have been used to study cellular processes~\cite{Lutkenhaus2007}, and general pattern emergence.~\cite{Vanag2009} 

A range of software have been developed for simulations of phase-field models. We will not review them, but some notable examples include MOOSE \cite{Permann2020}, PRISMS-PF~\cite{DeWitt2020}, FiPy~\cite{Guyer2009}, and OpenPhase~\cite{Tegeler2017}. These packages provide varying degrees of user configuration and flexibility in implementing phase-field models, including support for coupling with material properties.

Since the release of the original version of \SymPhas{} in 2022 \cite{Silber2022-pz}, two more software packages have appeared, {PhaseFieldX}~\cite{Castillon2025} and {MicroSim}~\cite{Dutta2025}. {PhaseFieldX} is an open-source package focused on phase-field fracture modeling built on {FEniCSx}~\cite{Baratta2023} that takes advantage of the FEniCSx high-level Python interface for domain and phase-field model definition as well as the built-in MPI~\cite{mpi40} parallelism in FEniCSx.
MicroSim, developed as part of the National Supercomputing Mission of India
is built upon selected features of the well-established OpenFOAM~\cite{Jasak2009} and AMReX~\cite{Zhang2021} solvers.
It also includes a GPU solver for the Kim-Kim-Suzuki (KKS)~\cite{Kim_1999} model using CUDA. Both PhaseFieldX and MicroSim can simulate phase-field models in 2- and 3-dimensions.

While GPU acceleration is available in some modern frameworks for instance, through backend libraries like PETSc in MOOSE or for specific models as in MicroSim, the direct compilation of high-level, symbolic model definitions into optimized GPU kernels remains a significant challenge. Many platforms that offer symbolic capabilities rely on runtime interpretation or code generation strategies that are not designed for the GPU architecture, while GPU-centric codes often require users to write low-level CUDA, sacrificing model development flexibility. This creates a gap for a framework that unifies high-level symbolic abstraction with high-performance, GPU-native execution.

The above limitations motivated the original development of \SymPhas{}, 
a simulation framework that implements compile-time symbolic algebra for defining phase-field, phase-field crystal, and reaction–diffusion models in 1-3 spatial dimensions. 
These features have made \SymPhas{} a practical tool for generating synthetic datasets used in machine learning with phase-field data~\cite{Kiyani2022a, Kiyani2023}.
While the original \SymPhas{} allowed users to write new models from their governing equations and adjust parameters via an external configuration file without recompilation, it was limited to single-node, shared-memory parallelism.

In this work, we present a significant update to the original \SymPhas{}. The new version, \SPnew{}, combines a more powerful symbolic algebra library with MPI parallelization and a high-performance, CUDA-based GPU-accelerated simulation framework that provides
user-specified control over the accuracy of finite difference stencils.
The symbolic library now includes directional and functional derivatives, symbolic summation, and integration, enabling users to define models directly from a free energy functional without manually deriving the equations of motion. Furthermore, mathematical expressions are compiled into optimized custom CUDA kernels, allowing simulations to be run entirely on the GPU and making \SPnew{} well-suited for modern high-performance computing environments.

\section{Software Functionalities}

The original \SymPhas{}~\cite{Silber2022-pz} is a modular C++ library designed for high-performance simulation of phase-field models consisting of a symbolic algebra library, numerical solver infrastructure, model abstraction classes, input/output management, configuration parsing, and utility functions. The architecture supports extensibility and customization by allowing users to write new simulation applications from scratch or extend existing examples. Each module can be used independently or integrated 
to suit 
specific requirements, including custom file formats or specialized solvers.

The primary feature set of the original \SymPhas{} is its symbolic algebra engine, which operates entirely at compile-time and offers an interface to manipulate, simplify, and transform mathematical expressions symbolically. This abstraction shifts focus from implementation details to 
model construction. It allows users to encode models at a mathematical level, avoiding boilerplate code and reducing implementation errors. The symbolic framework supports models involving multiple fields, arbitrary spatial dimensionality (1D, 2D, 3D), and a range of boundary conditions.

Numerical solvers in \SymPhas{} are written following object-oriented design principles, separated from the model definitions and algebraic formulation. When a phase-field model is defined, its equations are converted to symbolic partial differential equations (PDEs), which are then passed through a standardized application programming interface (API) to the solver classes. This decoupling enables advanced users to create custom solvers or modify existing ones with minimal effort.

\subsection{New Features in \SPnew{} }

First, \SPnew{} introduces major advancements in the symbolic algebra library. The library now supports pure symbolic terms, additional differentiation symbols including directional and functional derivatives, symbolic differentiation, symbolic summation and integration, and free energy–based model construction. The new functionalities enable simpler implementation of phase-field models from their mathematical formulation by removing the need for manually taking the  derivatives, and expand the scope of expressions that can be written. A particularly significant addition in the new version is the ability to use the GPU-accelerated CUDA backend for the finite difference solver, resulting in improvements in runtime of up to 1,000 times for large system sizes of some tested models.

\subsubsection{Symbolic Algebra Enhancements}

The symbolic algebra system in \SPnew{} has undergone a substantial extension relative to the original release, incorporating a broader range of expression types and supporting the construction of significantly more complex mathematical formulations. A key addition is the distinction between \textit{evaluable} and \textit{non-evaluable} expressions: the former can be reduced to a numerical value upon evaluation, while the latter represent abstract entities or structural placeholders within symbolic formulations. The non-evaluable category includes operators, and new types such as pure symbols and summation indices.

One of the key updates in \SPnew{} is the redesign of the symbolic object representing a phase-field variable to include subtypes that distinguish between statically and dynamically indexed order parameters, outlined in Table~\ref{table:expression-terms}. In the static case, the set of fields is known at compile time (e.g., a binary system), and the fields can be addressed directly by indexing with 0 or 1. In the dynamic case, the number of fields is supplied by the runtime configuration, and while direct indexing is valid, the equations also support generic indexing with an expression symbol (e.g. $i$). This mechanism allows a single model definition to be reused for multiphase systems with an arbitrary number of fields (e.g., multiple cells or layers) without rewriting the equations.
This capability is particularly useful for enabling reusable model definitions in which the number of fields is not hardcoded. For example, it allows users to define a model with an arbitrary number of coupled fields that follow the same free energy and coupling structure, without having to re-implement the model for each specific case.

The set of expression objects used to construct symbolic expression trees has been expanded and refined. Core components such as binary operators have been improved for greater consistency, and several new operator types have been introduced to support more sophisticated formulations. These are comprehensively listed in Table~\ref{table:expression-evaluable}.

To support compile-time arithmetic in symbolic transformations, constant expression types, such as rational literals, have been added and are described in Table~\ref{table:expression-constants}. These constants are crucial for enabling finite-difference stencils to be derived from a set of linear equations, and expressing normalized coefficients in symbolic rules. 

\SPnew{} also introduces the new literal type \lstinline|OpTensor| to support vector- and tensor-valued symbolic expressions. This object encodes individual components with explicit compile-time indices, enabling structured manipulation of vector and tensor values. Symbolic rules have been implemented for operations such as vector addition, dot products, and component-wise arithmetic. Expressions involving vector fields can be naturally decomposed and interpreted component-wise, allowing numerical solvers to be applied to each component in a consistent, rule-based way.

Much of the newly introduced functionality centers around non-evaluable types, which serve as placeholders for algebraic operations that are resolved through rule-based evaluation. This allows phase-field equations to be defined in a compact and intuitive form, deferring construction of the expression to symbolic rules that apply the appropriate substitutions and expansions. 
For phase-field models derived from a free energy functional, the most important new feature is the \lstinline{SymbolicFunctionalDerivative} type. This symbol acts as a placeholder that instructs the symbolic engine to automatically compute the variational derivative of a given free energy expression, which is the core mechanism that enables the automated derivation of the equations of motion. Conversely, for models where the equations of motion are given directly, such as in many reaction-diffusion problems, other non-evaluable types are critical. These include \lstinline{Symbol} types, which act as placeholders for parameters, and \lstinline{SymbolicSum} expressions, which are used to construct interactions in multi-component systems. These tools, combined with the library of symbolic derivative operators, allow for the direct and intuitive construction of the final equations. All the relevant expression types are listed in Tables~\ref{table:expression-derivatives} and~\ref{table:expression-symbols}.

\subsubsection{Pure Symbols and Symbolic Functions}

Pure symbols in \SPnew{} are implemented as template specializations of a base \lstinline|Symbol| class (see Table~\ref{table:expression-symbols}). These represent uniquely identifiable, non-evaluable entities that can be substituted and manipulated symbolically, independent of any specific variable or numeric value. Unlike expressions that are evaluable, symbols serve as placeholders to support symbolic substitution and identity-based simplification. 
For example, expressions such as $x + x = 2x$ or $x - x = 0$ rely on recognizing $x$ as the same symbolic entity. These symbols integrate with the broader term and operator hierarchy and are essential for parameterizing expressions, supporting algebraic transformations, and defining generic operations such as directional derivatives or symbolic integrals.

\subsubsection{Advanced Derivative Support}

Symbolic derivatives are supported to arbitrary order, including directional and mixed derivatives. \SPnew{} includes additional symbolic expression types to manage this functionality, which are listed in Table~\ref{table:expression-derivatives}. With finite-difference based solvers, the derivative symbols can directly call the corresponding stencil function, mapped by dimension, order and accuracy. \SPnew{} also enables the user to manually implement stencils when required when high-order derivatives demand better resolution or stability. The symbolic algebra system reduces compound expressions such as $\nabla\cdot\nabla\phi$ into a canonical form $\nabla^2 \phi$, avoiding redundant operator application and ensuring the correct stencil is used by the solver. This is implemented by expanding differential operators into vector components, performing symbolic dot products, and applying simplification rules for second derivatives.

C++ templates are used to implement type traits that can dispatch expressions based on whether the expression is evaluable or non-evaluable, and its dimensionality. Type traits are compile-time predicates and constants that attach properties to types, and are used in constrained overloads to select the appropriate transformation rules, stencils, and kernel specializations. The symbolic algebra rules apply recursively to ensure correctness of vector calculus identities and support complex nested expressions.

\subsubsection{Summation Expressions}

Summations can now be defined symbolically over fixed or dynamically sized field lists. Summation terms support index offsets (e.g., $v_{i+1}$), arbitrary bounds, and nested structures. The summand can include any symbolic expression and use placeholders for indices or variables. These symbolic summations are compiled into iteration logic that constructs the expression statically, checking at compile time that the bounds, indices, and terms are valid.

Summation support is particularly useful in constructing multi-phase-field models, where the free energy is defined over a set of fields. Nested sums allow interaction terms between fields to be written in a compact and expressive way. \SPnew{} ensures that the symbolic expressions generated through nested sums can be simplified and differentiated like any other symbolic object.

\subsubsection{Free Energy–Based Model Definitions}

One of the major features in \SPnew{} is the ability to define a model directly from its free energy functional: Users provide the free energy as an integral over a symbolic expression, and the software then automatically computes the variational derivative with respect to each order parameter field using symbolic rules, applies simplification, and substitutes the result into the dynamical equations.
This eliminates the need for manually deriving and coding the equations of motion, allowing faster model prototyping and reducing the potential for error. In general, the functional derivative is computed using the identity~\cite{Gelfand2000}:
\begin{equation}
\frac{\delta F}{\delta \phi} = \sum_{k=0}^{n} (-1)^k \nabla^{(k)} \left( \frac{\partial f}{\partial (\nabla^{(k)} \phi)} \right)
\label{eq:functional_derivative_general},
\end{equation}
where the functional $F$ is defined as
$$
F[\phi] = \int f\left( \phi, \nabla \phi, \nabla^2 \phi, \dots, \nabla^n \phi \right)  d^d x.
$$
Here, $f$  depends on the field $\phi$ and its derivatives up to order $n$. The notation $\nabla^{(k)} \phi$ denotes the $k^\mathrm{th}$ order derivative of $\phi$. If $F$ depends only on the first order derivative, Eq.~\ref{eq:functional_derivative_general} reduces to the familiar form
\begin{equation}
\frac{\delta F}{\delta \phi} = \frac{\partial f}{\partial \phi} - \nabla \cdot \left( \frac{\partial f}{\partial (\nabla \phi)} \right).
\end{equation}
In the current implementation, functional differentiation is implemented via compile-time symbolic differentiation of the integrand, and is limited to the case where the free energy depends on the first order derivative. This limitation will be addressed in a future release.

The free energy definition and associated dynamics are passed using C++ macros, which expand to type-safe class definitions at compile time. Various samples of model definitions are provided in the Github repository.

\subsubsection{Compile-Time Stencil Generation}

\SPnew{} computes finite difference stencil coefficients at compile time by solving a linear system derived from polynomial interpolation~\cite{Patra2006}. For a derivative operator of order $p$ and desired accuracy $n$, a polynomial of order $q = p + n - 1$ is constructed, and the coefficients are solved such that the operator is approximated to $\mathcal{O}(h^n)$ accuracy.

Stencil shapes are simplified using symmetry rules that reduce the number of unknowns~\cite{Patra2006} to provide better compilation times. The derived stencil coefficients are encoded as a new C++ type and built into the solver implementation to eliminate any runtime overhead. This design supports arbitrary derivative types and orders across 1D, 2D, and 3D grids, offering high accuracy without requiring users to manually define or verify stencil coefficients.

Together, these features in \SPnew{} form a highly expressive, efficient, and extensible platform for developing and simulating complex phase-field models with minimal manual intervention.

\subsubsection{Construction of Semi-Implicit Fourier Spectral Scheme}

\SPnew{} supports the construction of spectral solver schemes~\cite{Chen1998, Yoon2020} using template functions that implement rule-based logic and operate directly on the structure of the symbolic expression tree to parse and transform the governing equations of a phase-field or a reaction-diffusion model. This functionality identifies differential operators, distinguishes linear from nonlinear terms, and reorganizes the expression into a form amenable to spectral integration, typically separating linear terms evaluated in Fourier space from nonlinear terms evolved in real space. While this capability was present in the original version of \SymPhas{}, it has been significantly extended in \SPnew{} through enhancements to the symbolic algebra framework, enabling more robust expression analysis and transformation.

\subsection{CUDA Implementation}

CUDA is a programming model developed by NVIDIA for writing general-purpose computations on GPUs~\cite{Nickolls2008-yf}.
It enables direct control over GPU memory, thread hierarchy, and synchronization, allowing parallel algorithms to be expressed efficiently on hardware with thousands of concurrent execution units. CUDA is widely used in scientific computing
due to its ability to execute fine-grained, data-parallel computations with low memory latency and high arithmetic throughput~\cite{Pandey2022-ye}.

In \SPnew{}, a complete GPU backend based on CUDA has been implemented, enabling symbolic algebra-defined numerical schemes to run efficiently on GPU architectures. To facilitate this, symbolic expressions -- originally represented as C++ template-based expression trees -- are reconstructed using a corresponding set of CUDA-specific symbolic classes. This results in an equivalent expression tree tailored for device execution. The CUDA-specialized tree is then passed as a template argument to a generic kernel function, allowing the compiler to generate an optimized kernel for the specific numerical scheme encoded in the symbolic expression.

GPU kernels are implemented as templated C++ functions, into which the symbolic expressions are substituted at compile time. Each expression becomes a specialized function kernel tailored to the specific numerical scheme, dimensionality, and field type. We define templates for scalar, vector, and complex fields
in 1-3 spatial dimensions. This design provides users with CUDA kernels without the need of manual GPU programming.

In GPU-based simulation, all field data, temporaries, and evaluation buffers reside on the device throughout the simulation. Data is transferred back to the host only when writing output to disk, minimizing host-device communication overhead. This design ensures that large-scale simulations can fully leverage GPU memory bandwidth and parallel throughput.

The flexibility of writing symbolic mathematical expressions on the host is preserved in the GPU backend. Users may construct arbitrary expressions with all the flexibility presented in the previous sections, 
and the symbolic library backend will convert the expressions into efficient CUDA kernels. This same infrastructure can be extended in future releases to GPU-accelerated spectral solvers, enabling symbolic Fast Fourier Transform (FFT)-based models such as the spectral solver that is currently implemented only on the host, to benefit from the same compile-time expression specialization.

\section{Performance}

To evaluate the performance improvements provided by GPU acceleration in \SPnew{}, we benchmarked four representative phase-field models: Models A, B, C, and H of the Hohenberg-Halperin classification~\cite{Hohenberg1977}
spanning both conserved and non-conserved dynamics using single order-parameter phase fields, and coupled two order-parameter scalar and vector order parameters. In the equations below (defined with each model), the constants have been set to unity.

Model A corresponds to non-conserved order parameter dynamics and is known as the Allen–Cahn model~\cite{Allen1975}, describing chemical potential driven field relaxation. Model B implements conserved dynamics, described by the Cahn–Hilliard equation~\cite{Cahn1958} and typically characterizes spinodal decomposition. Model C couples the dynamics of a non-conserved order parameter to a conserved scalar field, a scenario commonly used to represent interactions between structural and concentration fields~\cite{Castillo2015}. Model H describes a conserved scalar order parameter coupled to a convective vector field, which can represent, for example, current-driven spinodal decomposition~\cite{Farrell1989}.

\subsection{Implementations}

In the original release of \SymPhas{}, model implementations were specified explicitly in terms of their evolution equations. In \SPnew{}, the new symbolic expression library allows models to be defined directly from their free energy functionals, from which dynamical equations are automatically derived using symbolic functional differentiation.

Below is a brief summary of the macros and expressions used in the definitions of the above models. These are also available at the GitHub site:
\begin{itemize}
    \item \lstinline{MODEL}: Declares a new model. The first argument is the model name.
    \item \lstinline{FREE_ENERGY}: Defines a model using a free energy expression. Takes the dynamics type, free energy expression, and order parameters as arguments.
    \item \lstinline{EQUATION_OF}: Used when the evolution equations must be manually derived from the free energy functional (e.g., for coupled systems).
\end{itemize}

Field and expression macros include:
\begin{itemize}
    \item \lstinline{SCALAR}, \lstinline{VECTOR}: Specify the type of each order parameter field.
    \item \lstinline{INT}: Represents a symbolic integral.
    \item \lstinline{SUM}: Represents a symbolic summation over an index (e.g., \lstinline{ii}).
    \item \lstinline{op}, \lstinline{op_ii}: Order parameter fields.
    \item \lstinline{c}: Model constants defined in the configuration.
    \item \lstinline{LANDAU_FE}: Shorthand for the polynomial Landau free energy density.
    \item \lstinline{lap}, \lstinline{grad}: Laplacian and gradient differential operators.
\end{itemize}

\paragraph{Model A}  
This model describes a non-conserved scalar order parameter evolving under Allen–Cahn~\cite{Allen1975} dynamics:
\begin{equation}
\frac{\partial \phi}{\partial t} = \nabla^2 \phi + c_1 \phi - c_2 \phi^3\,.
\label{eq:model-a}
\end{equation}
The symbolic implementation uses the \lstinline{FREE_ENERGY} macro to define the system from a Landau free energy~\cite{Kardar2013} (\lstinline{LANDAU_FE}) in the form of a functional. The order parameter is a scalar field, declared as \lstinline{SCALAR}, and the dynamical class is \lstinline{NONCONSERVED}.
\begin{lstlisting}
#define phi op(1)
MODEL(MA_FE, (SCALAR),
  FREE_ENERGY((NONCONSERVED), 
    INT(LANDAU_FE(phi, c(1), c(2))))
)
\end{lstlisting}
The term \lstinline{op} refers to the order parameter, and \lstinline{c} refers to the configuration-provided coefficients corresponding to the coefficients $c_1$ and $c_2$ in Equation~\ref{eq:model-a} above. 

\paragraph{Model B}  
This model implements conserved scalar dynamics, and is also known as the Cahn–Hilliard~\cite{Cahn1958} equation:
\begin{gather}
\frac{\partial \rho}{\partial t} = \nabla^2 \left( -\nabla^2 \rho - c_1 \rho + c_2 \rho^3 \right)
\label{eq:model-b}
\end{gather}
The structure is nearly identical to Model A, but the dynamics are set to \lstinline{CONSERVED}:

\begin{lstlisting}
#define rho op(1)
MODEL(MB_FE, (SCALAR),
  FREE_ENERGY((CONSERVED), 
    INT(LANDAU_FE(rho, c(1), c(2))))
)
\end{lstlisting}

\paragraph{Model C}  

%



This model couples a non-conserved scalar order parameter to a conserved density~\cite{Hohenberg1977}:
\begin{gather}
\frac{\partial \phi}{\partial t} = \nabla^2 \phi + c_1\phi - c_2\phi^3  - 2\phi \rho\,,\\
\frac{\partial \rho}{\partial t} = \nabla^2 \rho + \nabla^2 (c_3 \phi^2)
\label{eq:model-c}
\end{gather}
The free energy functional, including a coupling term, is conveniently defined:

\begin{lstlisting}
#define phi op(1)
#define rho op(2)
MODEL(MC_FE, (SCALAR, SCALAR),
  FREE_ENERGY(
    (NONCONSERVED, CONSERVED), 
    INT(LANDAU_FE(phi, c(1), c(2)) + 0.5_n * rho * rho 
        - c(3) * phi * phi * rho)
  )
)
\end{lstlisting}
We have set $c_3$ (corresponding to \lstinline{c(3)}) to 0.5 for the simulations. The value \lstinline{0.5_n} is a constant in the symbolic algebra.

\paragraph{Model H}  

%
%
%
%
%

Model H describes a conserved scalar field coupled to a convective vector field~\cite{Farrell1989}:
\begin{equation}
\begin{aligned}
\frac{\partial m}{\partial t} &= \nabla^2 \left( -\nabla^2 m - c_1 m  + c_2 m^3\right) - c_3 \nabla\cdot \left( m \mathbf{j} \right) , \\
\frac{\partial \mathbf{j}}{\partial t} &= \Pi^{\perp}\left( \nabla^2 \mathbf{j} - c_3 m \nabla \left( \nabla^2 m + c_1 m - c_2 m^3  \right)\right) \,.
\end{aligned}
\label{eq:model-h}
\end{equation}

Typically, Model H is defined with a transverse operator applied to the dynamical equation for $\mathbf{j}$ as shown above, but for the purposes of this work, particularly since we are comparing the simulation performance, we set $\Pi^{\perp}$ to unity.

Since this system requires a more specialized evolution equation, the \lstinline{EQUATION_OF} macro is used to explicitly construct the dynamical terms from the variational derivatives:

\begin{lstlisting}
#define m op(1)
#define F_m DF(1)
#define j op(2)
#define F_j DF(2)
MODEL(MH_FE, (SCALAR, VECTOR),
  FREE_ENERGY(
    (
      EQUATION_OF(1)(-lap(-F_m) - c(3) * grad * (m * F_j)),
      EQUATION_OF(2)(lap(F_j) + c(3) * m * grad(-F_m))
    ),
    INT(LANDAU_FE(m, c(1), c(2)) + 0.5_n * j * j)
  )
)
\end{lstlisting}

\paragraph{Symbolic constructs.}  
All models share a common set of macros and symbolic terms:
\begin{itemize}
  \item \lstinline{MODEL}, \lstinline{FREE_ENERGY}, and \lstinline{EQUATION_OF} are used to declare models and specify dynamical equations.
  \item Field types are specified with \lstinline{SCALAR} and \lstinline{VECTOR}.
  \item \lstinline{INT}, \lstinline{SUM} denote integrals and index summations expressions.
  \item \lstinline{LANDAU_FE} is a shorthand for the polynomial Landau free energy density.
  \item \lstinline{op}, \lstinline{op_ii} refer to order parameter fields; \lstinline{c(n)} are user-provided constants from the configuration file.
  \item \lstinline{grad} and \lstinline{lap} denote gradient and Laplacian operators; \lstinline{DF(n)} denotes the functional derivative with respect to field $n$.
\end{itemize}
This symbolic formulation abstracts away the numerical details of model implementation, and allows models to be flexibly assembled, modified, and extended within the same infrastructure. It also supports hybrid symbolic-numeric schemes such as manually defining equations from a free energy, as done in Model H.

\subsection{Benchmarks}

Runtimes of simulations on both CPU and CUDA backends using the finite difference solver were recorded for a range of system sizes in 2 and 3 dimensions. System specifications are detailed in Table~\ref{tab:benchmark-systems}.
The benchmark is set to $1,000$ time steps on square grids with periodic boundary conditions and side lengths ranging from $64$ at the lower end, to $8,192$ for 2D or to $1,024$ for 3D at the upper end, depending on the maximum memory available on the system (either RAM or VRAM). Model A uses a timestep of 0.1, while Models B, C, and H use a timestep of 0.01. All tests were performed with double precision arithmetic. The memory footprint for each model in units of bytes, given as 8 times the system size (since the double precision word length is 8 bytes), is shown in Table~\ref{tab:memory-footprint}. If single precision is used, which has a word length of 4 bytes, the memory footprint is halved, and the maximum possible system size would be doubled. In the benchmarks, runtime was averaged over five independent simulations per system size initially starting with uniformly distributed random noise.
We included the Nvidia H100 GPU (Hopper architecture) in our performance tests, which is the newest GPU available on the recently commissioned ``nibi'' Digital Research Alliance of Canada~\cite{AllianceNibi} cluster.
\begin{table}[tb]
    \centering
    \begin{tabular}{|c|c|l|l|}
        \hline
        \textbf{System} & \textbf{Type} & \textbf{Model} & \textbf{Memory} \\ \hline
        Desktop & CPU & \footnotesize {Intel(R) Core(TM) i7-9700 8-Core @ 3.00GHz} & 32GB \\
                 & GPU & \footnotesize{GeForce RTX 2070 SUPER} (Turing architecture) & 8GB \\ \hline
        Desktop & CPU & \footnotesize{AMD Ryzen 7 5800 8-Core @ 3.40GHz} & 32GB \\
                 & GPU & \footnotesize{GeForce RTX 3080} (Ampere) & 12GB \\ \hline
        Laptop & CPU & \footnotesize{Intel(R) Core(TM) i7-1195G7 4-Core @ 2.90GHz} & 16GB \\
                 & GPU & \footnotesize{GeForce GTX 1650} (Turing) & 4GB \\ \hline
        Cluster & GPU & \footnotesize{H100 Tensor Core SXM (Hopper)} & 80GB \\ \hline
    \end{tabular}
    \caption{System specifications used in benchmarking. The first three devices on desktop and laptop systems are on private devices.
    The H100 GPU is on a compute node in the Compute Canada "nibi" cluster~\cite{AllianceNibi}. All simulations use a single GPU.}
    \label{tab:benchmark-systems}
\end{table}

\begin{figure}[tb]
    \centering
%
       \subfloat[Model A]{
        \includegraphics[width=0.4\textwidth]{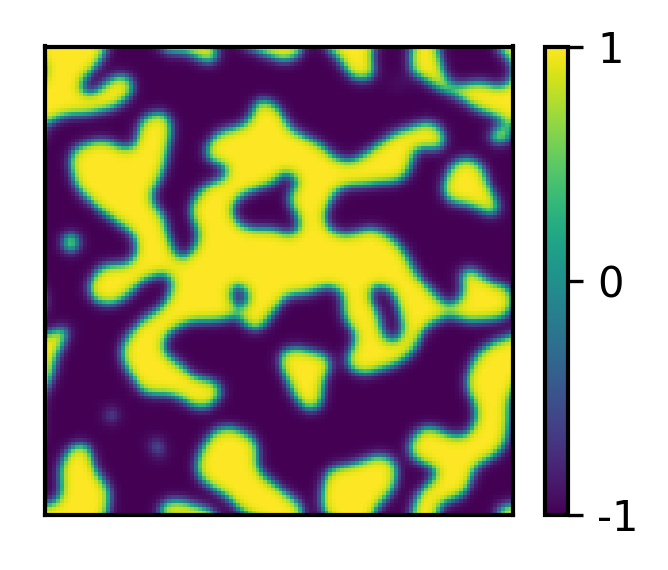}        
        \label{fig:modela-2d}
        }
    \hfill
       \subfloat[Model B]{
       \includegraphics[width=0.4\textwidth]{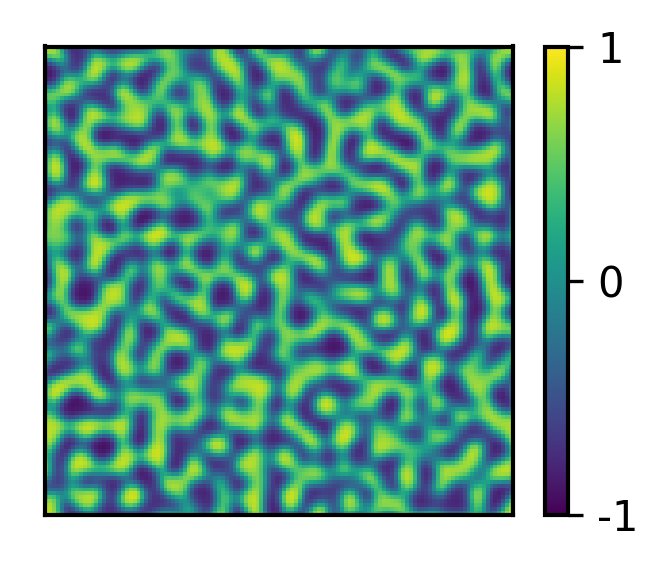}
        \label{fig:modelb-2d}
        }

    \vspace{1em}

       \subfloat[Model C: Non-conserved Field]{
       \includegraphics[width=0.4\textwidth]{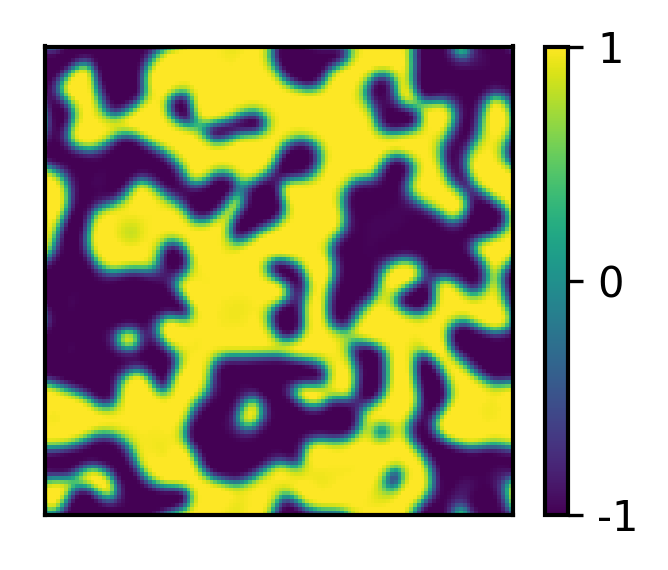}       
        \label{fig:modelc-2d-f1}
        }
    \hfill
       \subfloat[Model C: Conserved Field]{
       \includegraphics[width=0.4\textwidth]{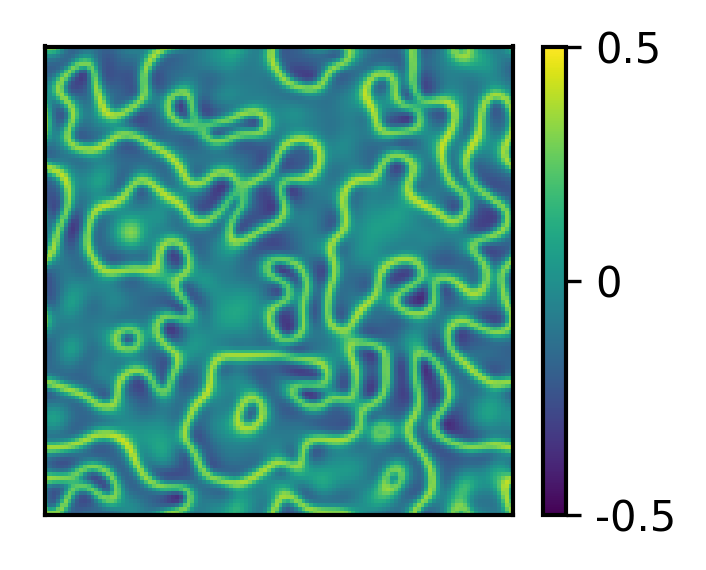}
        \label{fig:modelc-2d-f2}
        }

    \vspace{1em}

       \subfloat[Model H: Scalar Field]{
       \includegraphics[width=0.4\textwidth]{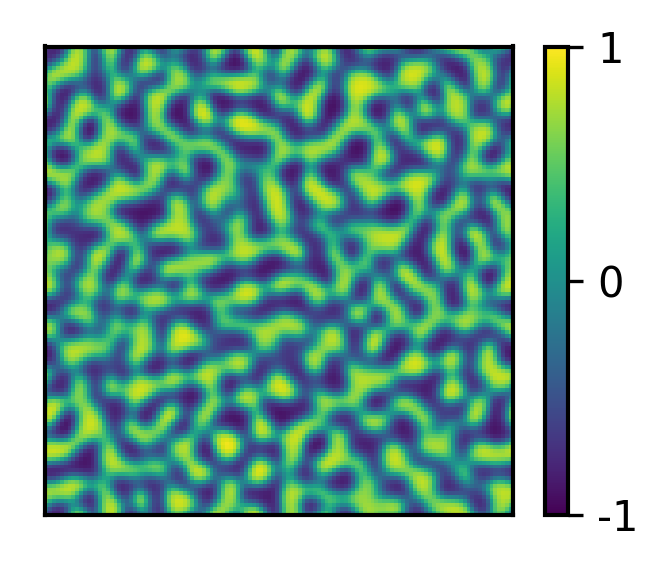}       
        \label{fig:modelh-2d-f1}
        }
    \hfill
      \subfloat[Model H: Vector Field]{
       \includegraphics[width=0.4\textwidth]{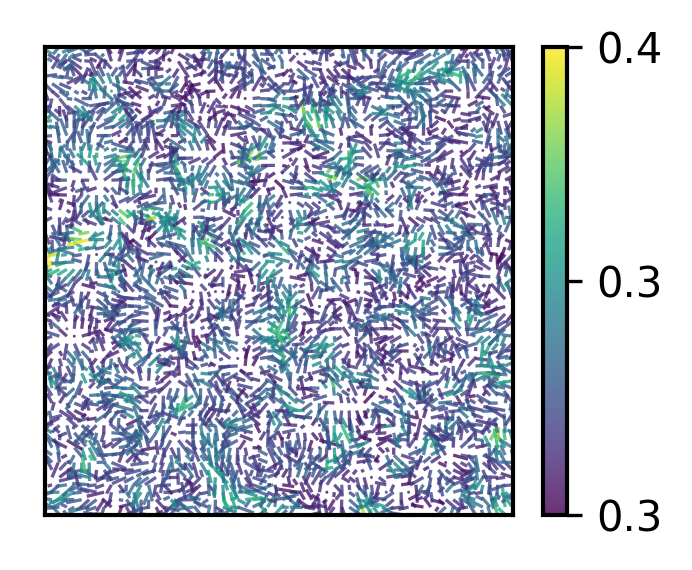}      
        \label{fig:modelh-2d-f2}
        }
    \caption{Two-dimensional simulation results for Models A, B, C, and H on grids of $128 \times 128$ with periodic boundaries. Models C and H involve two coupled fields and are displayed side-by-side. Each simulation was run for 1,000 iterations. }
    \label{fig:all-models-2d}
\end{figure}

\begin{figure}[tb]
    \centering
       \subfloat[Model A]{
        \includegraphics[width=0.4\textwidth]{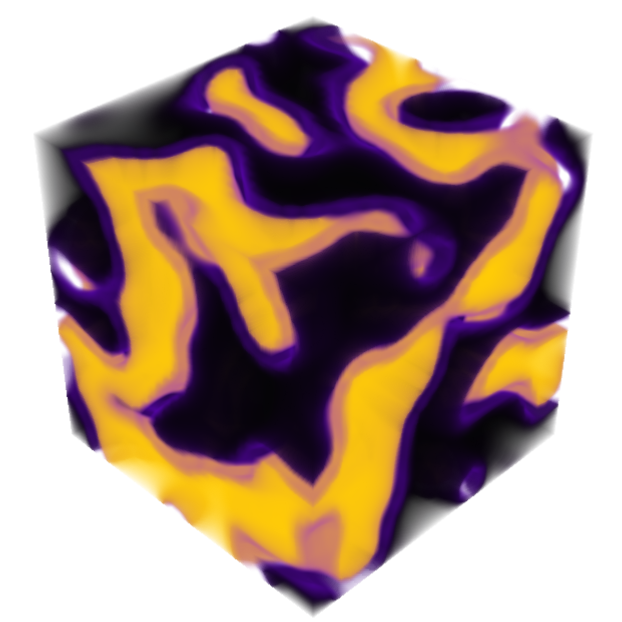}
        \label{fig:modela-3d}
        }
    \hfill
        \subfloat[Model B]{
        \includegraphics[width=0.4\textwidth]{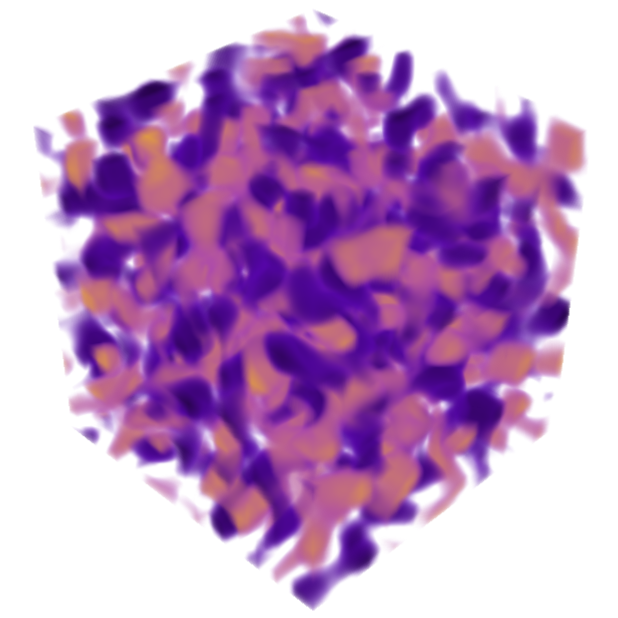}
        \label{fig:modelb-3d}
        }

    \caption{Visualizations from 3D simulations of Models A and B on a grid of $64\times 64\times 64$ with periodic boundaries. Each simulation was run for 1,000 iterations.}
    \label{fig:all-models-3d}
\end{figure}

\begin{figure}[tb]
    \centering
     \includegraphics[width=\textwidth]{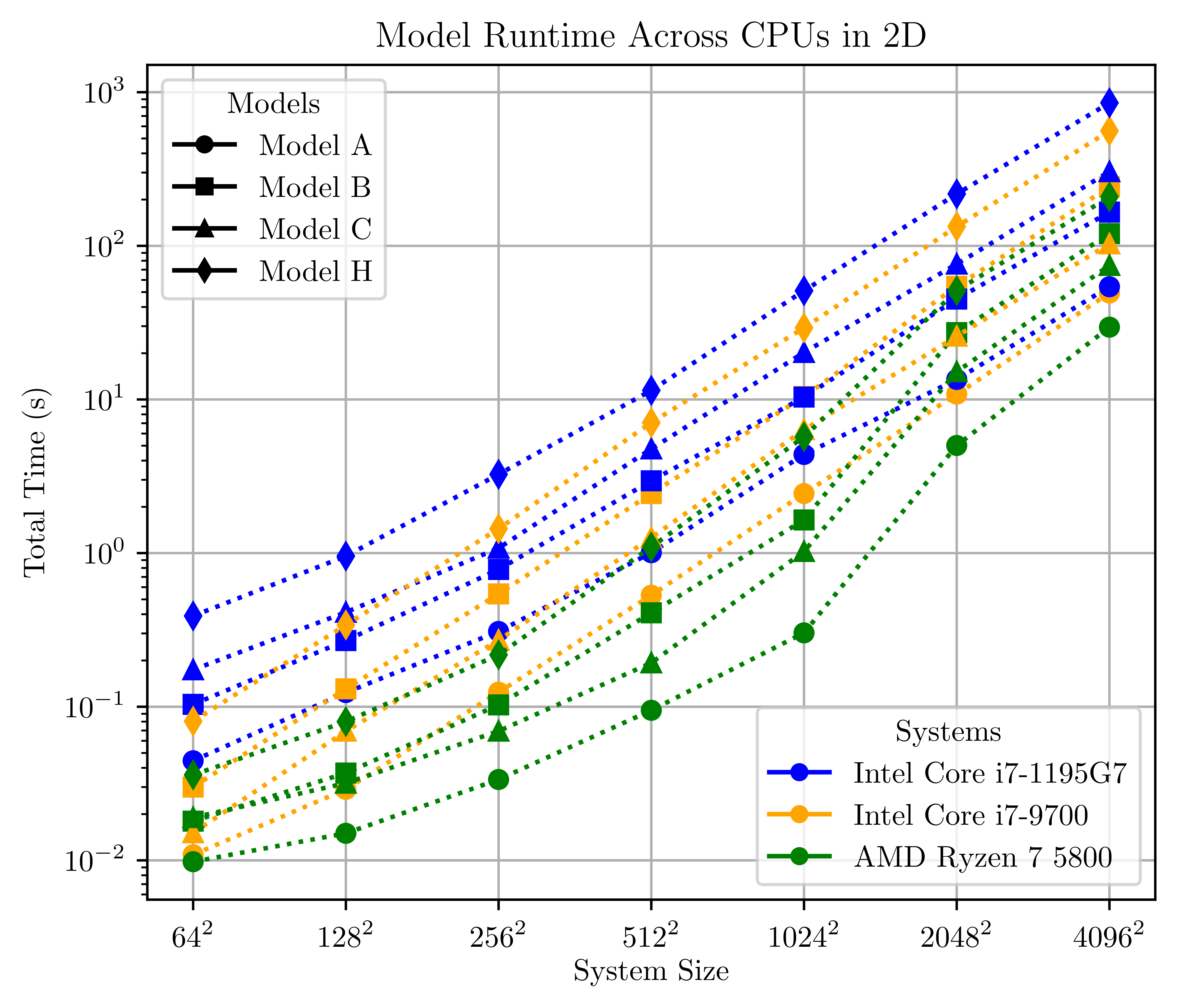}
    \caption{Simulation runtime of different CPUs using parallelization; for each system size, the runtime is averaged across 5 independent runs executed for 1,000 iterations. The scaling is close to 1 for all three different systems that were used.}
    \label{fig:cpu-peformance-2d}
\end{figure}

\begin{figure}[tb]
    \centering
     \includegraphics[width=\textwidth]{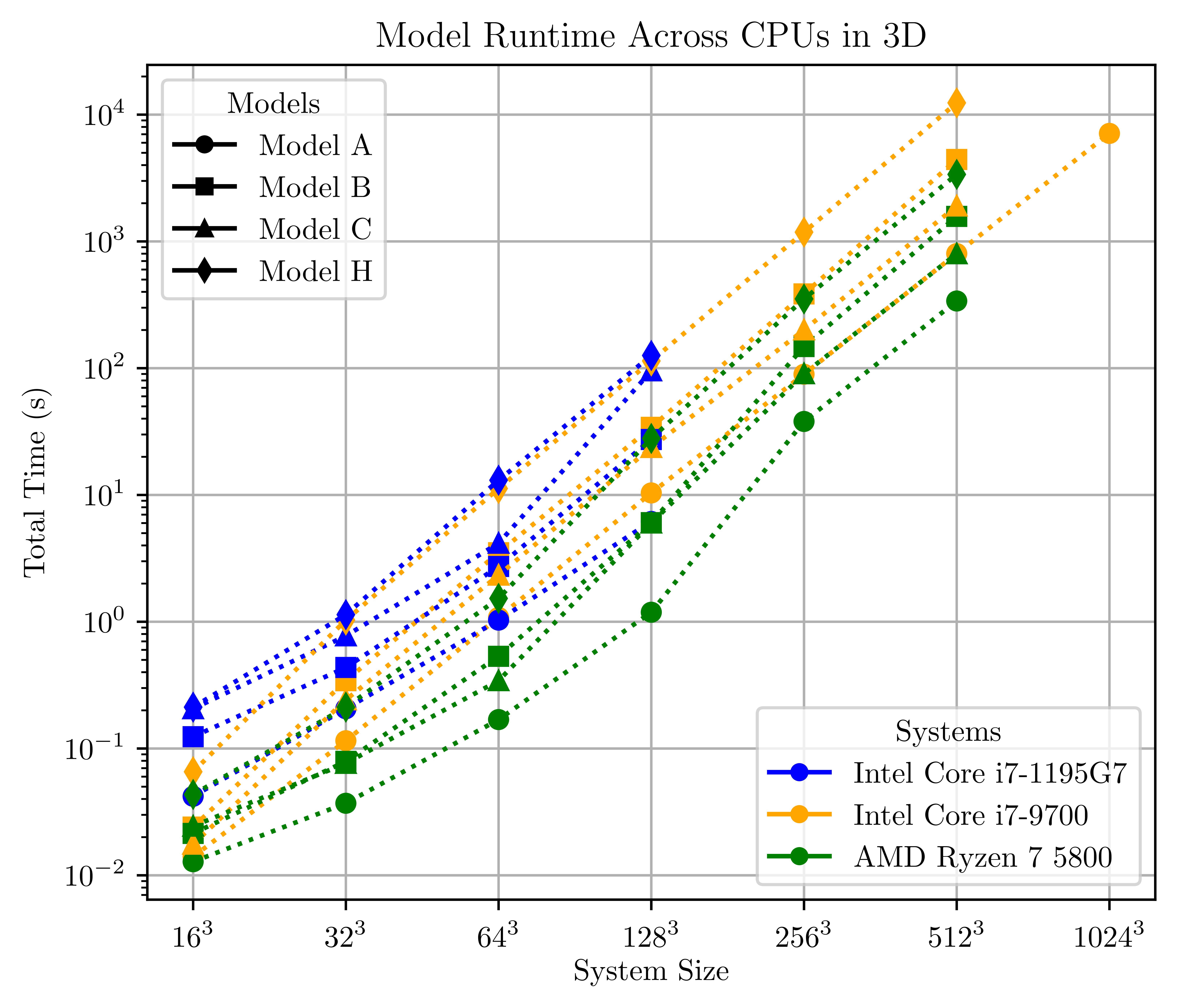}
    \caption{Simulation runtime of different CPUs using parallelization; for each system size, the runtime is averaged across 5 independent runs executed for 1,000 iterations. The scaling is close to 1 for all three different systems that were used. The models were simulated up to the maximum system size allowed by the available memory.}
    \label{fig:cpu-peformance-3d}
\end{figure}

\begin{figure}[tb]
    \centering
     \includegraphics[width=\textwidth]{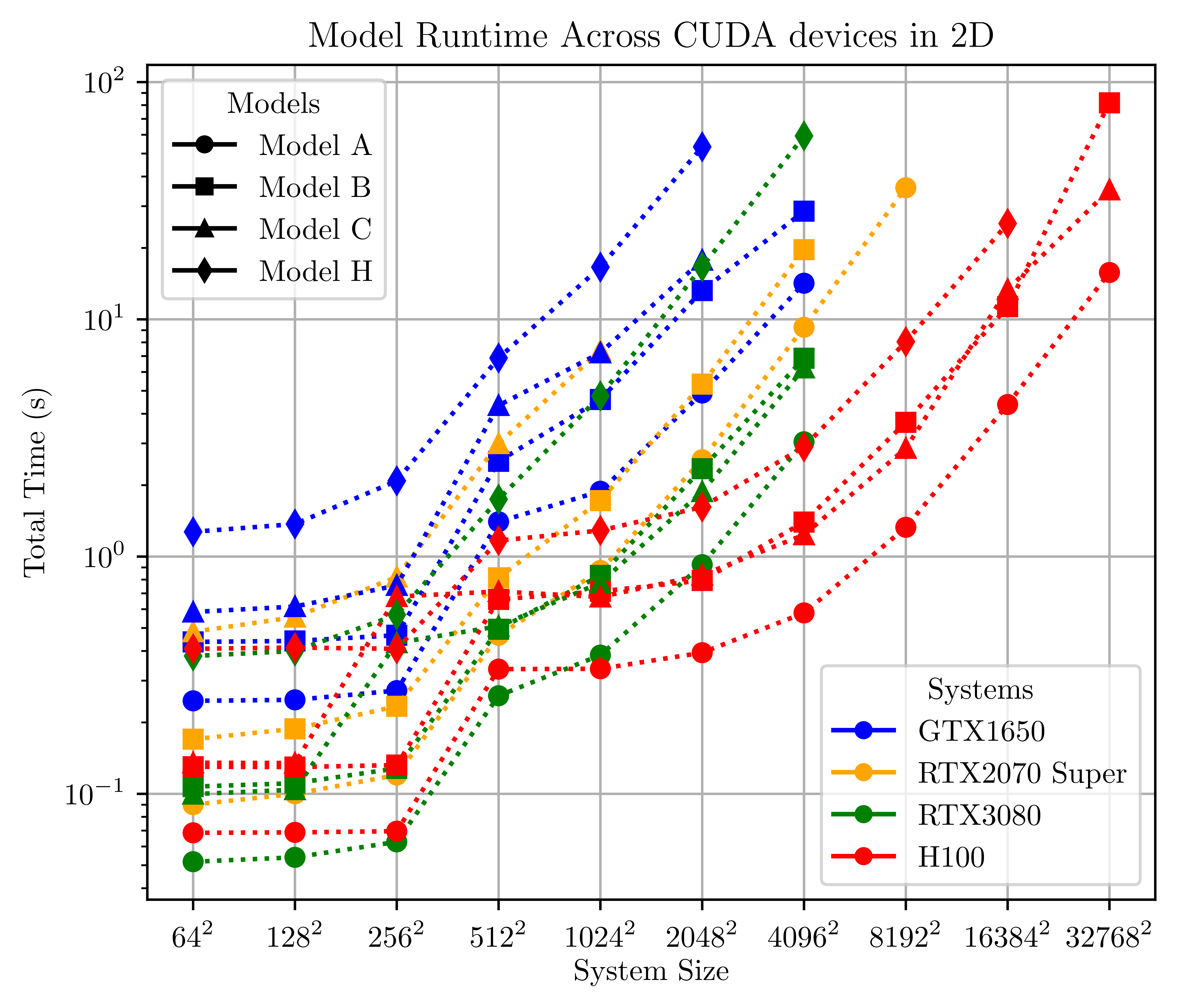}
    \caption{Simulation runtime of various CUDA-enabled GPUs; for each system size, the runtime is averaged across 5 independent runs executed for 1,000 iterations. The inconsistent scaling is due to caching and communication bottlenecks at smaller system sizes. The models were simulated up to the maximum system size allowed by the available memory on the GPU.
     }
    \label{fig:gpu-peformance-2d}
\end{figure}

\begin{figure}[tb]
    \centering
     \includegraphics[width=\textwidth]{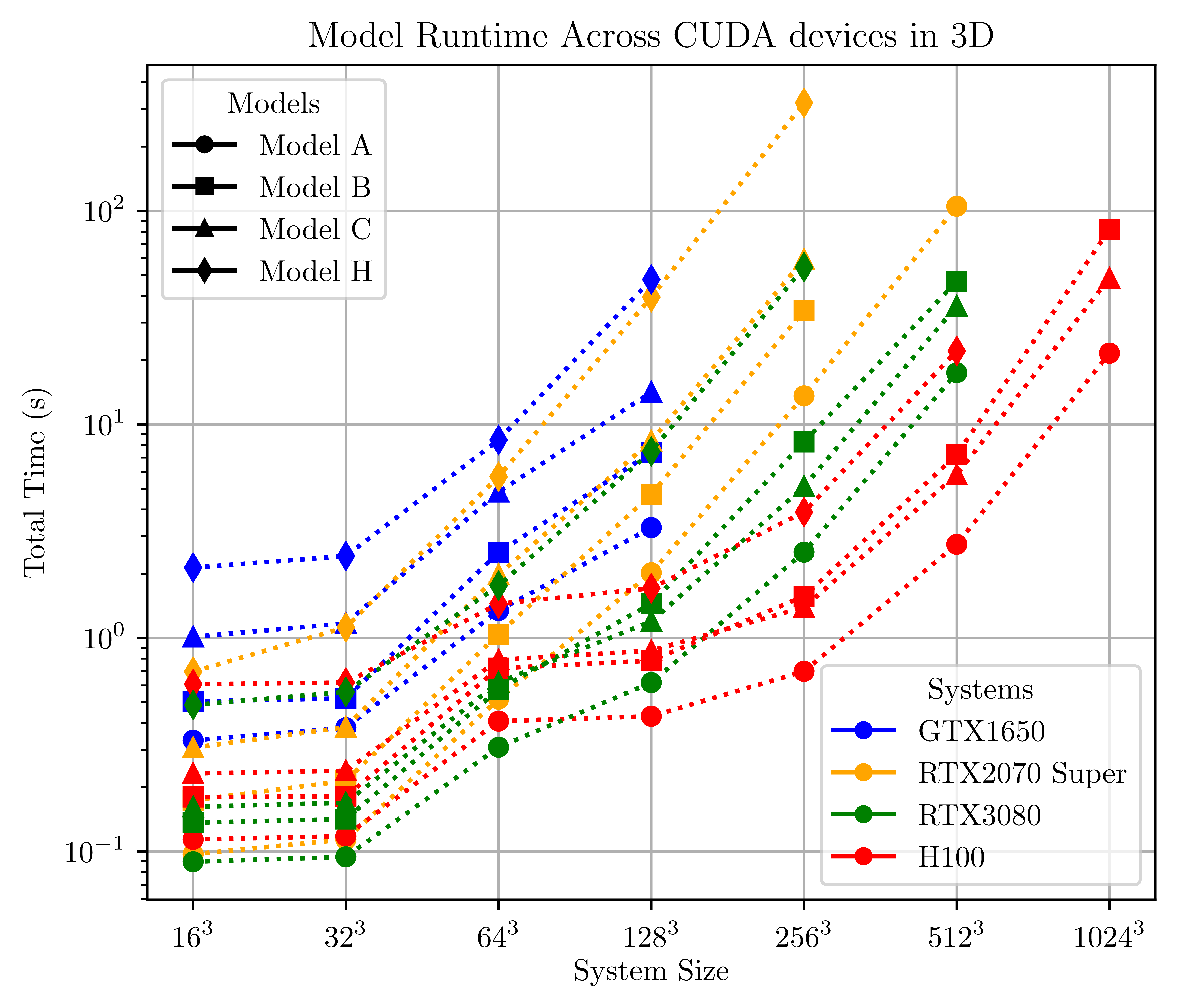}
    \caption{Simulation runtime of various CUDA-enabled GPUs; for each system size, the runtime is averaged across 5 independent runs executed for 1,000 iterations. The inconsistent scaling is due to caching and communication bottlenecks at smaller system sizes. The models were simulated up to the maximum system size allowed by the available memory on the GPU.}
    \label{fig:gpu-peformance-3d}
\end{figure}

We first present visualizations from sample simulations of models A, B, C and H in 2D, shown in Figure~\ref{fig:all-models-2d} and for Models A and B in Figure~\ref{fig:all-models-3d} for 3D.
The runtime results for the simulations of the four phase-field models using a CPU backend are plotted in Figure~\ref{fig:cpu-peformance-2d} in the 2D case, and in Figure~\ref{fig:cpu-peformance-3d} for the 3D case.
The results of the four different CUDA-enabled GPUs for 2D and 3D are plotted, respectively, in Figures~\ref{fig:gpu-peformance-2d} and~\ref{fig:gpu-peformance-3d}. 
No errors bars are visible since the runtimes are very close together. The system size was extended until the power scaling approaches 1, or farther if the system size was still far from the memory limit.
A direct comparison between GPU and CPU performance is not included, as differences in hardware architecture, memory bandwidth, and parallelization strategies make it difficult to establish a fair and representative baseline. 

For all models on CPU, the runtime scales very closely to $\mathcal{O}(N)$ with the number of grid points $N$. For CUDA, we observe an initial plateau in runtime across small system sizes, followed by a gradual increase in the runtime scaling as the system size grows. A notable discontinuous jump is also present once the system exceeds a critical threshold in size. The initial flat scaling regime is attributed to device synchronization overhead dominating the total runtime. During this stage, kernel launch latency and device-host coordination are the primary bottlenecks, rather than arithmetic throughput. This effect is particularly evident in the similar runtimes observed for both consumer- and data-center-class devices, such as the RTX~3080 and H100, respectively, despite their substantial differences in capability.

The influence of memory synchronization is also apparent when comparing the different phase-field models. Model A, which involves only a single scalar field, exhibits the shortest runtime in the flat regime. In contrast, Models B and C have an extra GPU-synchronization step for additional temporary fields that hold the result of an intermediate expression, prior to the solver iteration. This is for the $\nabla^2 (\phi^3)$ term that appears in the conserved dynamics, which requires that the nonlinear expression $\phi^3$ is evaluated and stored before the Laplacian operator is applied. This approach trades additional memory overhead and synchronization cost for improved performance by avoiding redundant computations during stencil evaluation. 
Model H, which involves $d+1$ fields in $d$ dimensions (a scalar field and a convective vector field), along with one temporary field, exhibits an even higher overhead due to increased synchronization demands. Its runtime exceeds that of Models B and C by approximately 30\% in 2D.

The observed jump in runtime beyond a certain system size likely corresponds to hardware-specific memory caching thresholds. Once the working set exceeds the available shared memory or L2 cache, the device must rely on slower global memory accesses, introducing latency that significantly impacts performance. This threshold depends on the number of fields, temporary arrays, and dimensionality, as they jointly determine the total memory footprint per grid point. In this regime, the performance becomes more sensitive to the memory bandwidth and architectural limitations of each GPU.

\begin{table}[tb]
    \centering
    \begin{tabular}{|l|l|l|l|}
        \hline
        \textbf{\small Model} & \textbf{$f$} & \textbf{$f_0$} & \textbf{$m$} \\ \hline
        Model A & 1 & 0 & 2 \\
        Model B & 1 & 1 & 3 \\
        Model C & 2 & 1 & 5 \\
        Model H & $1+d$ & 1 & $3+2d$ \\ \hline
    \end{tabular}
    \caption{
    Breakdown of the approximate on-device memory consumption for each model using double precision. The total number of bytes ($m$) required on the GPU is $2f + f_0$, which is the sum of the number of fields of the model ($f$), an equal number of solver buffers used during iteration, and any additional temporary arrays allocated by the expressions ($f_0$). The total memory footprint in bytes can be estimated by the formula Memory = $m\times N\times P$, where $N$ is the total number of grid points and $P$ is the precision size in bytes (8 for double, 4 for single).
    }
    \label{tab:memory-footprint}
    
\end{table}

\clearpage

\section{Conclusions}

\SPnew{} represents a significant advancement in capability compared to the previous version and further distinguishes it as a unique software among phase-field simulation frameworks. Users can study a wide range of applications in materials science, pattern formation, and reaction-diffusion systems, extending to sophisticated models with scalar-, vector- and complex-valued order parameters. The symbolic algebra library that supports this functionality has been reinforced with a larger inventory of symbols and options for defining phase-field models, notably with the ability to write free energy functionals and summations over arbitrary numbers of order parameters. Any model definition is automatically converted to a numerical scheme that is optimized by the compile-time expression trees, and can be automatically converted into CUDA kernels to run on GPUs.

The new CUDA implementation makes large-scale simulations practical and accessible, running efficiently even on consumer-grade NVIDIA GPUs. 
The performance benchmarks in this work were conducted using double precision arithmetic to ensure high numerical accuracy. The framework also supports single precision, which would halve the memory footprint and effectively double the maximum system size that can be simulated on a given GPU.
Our performance benchmarks demonstrate speedups of up to $\sim1,000\times$ compared to the original multi-threaded CPU version for large systems. This high-speed performance is particularly pronounced for the large system sizes where the computational workload surpasses the initial device synchronization overhead, allowing the simulations to fully leverage the parallel architecture.


\section*{Data availability} 

\SymPhas{} is open source and available at \\
\texttt{https://github.com/SoftSimu/SymPhas}

\section*{Conflict of Interest} 

None to declare.


\section*{Acknowledgments}

M.K. thanks  the  Natural  Sciences  and  Engineering  Research  Council  of  Canada (NSERC), Canada Research Chairs Program, and Foundation PS for financial support.  S.A.S. thanks NSERC for financial support through the Postgraduate Scholarships (PGS D) program
The Digital Research Alliance of Canada is thanked for  computational resources.


\newpage
\appendix

\setcounter{table}{0}
\renewcommand{\thetable}{A\arabic{table}}

\section{Symbolic Algebra Objects}
\label{app:symbolic-algebra-objects}
%

\begin{table}[h]
\centering
\caption{Objects used in constructing and representing a phase-field variable. These are not evaluated directly and are instead primarily used to implement rules of algebra between, or hold information about, phase-field variables. The columns v1 and v2 refer to the original version of \SymPhas{} and \SPnew{}, respectively.}
\small
\label{table:expression-terms}
\begin{tabular}{|p{0.3\linewidth}|p{0.5\linewidth}|l|l|}
\hline
\textbf{Symbol / Node} & \textbf{Description} & v1 & v2 \\
\hline
\texttt{NamedData<T>} & Represents a specific variable in a mathematical expression, primarily for implementing symbolic algebra rules such as simplification.&$\checkmark$&$\checkmark$ \\
\texttt{Variable<Z, T | NamedData>} & Represents a specific variable \texttt{T}, which can be a \texttt{NamedData}, in a mathematical expression, primarily for implementing symbolic algebra rules of simplification.&$\checkmark$&$\checkmark$ \\
\texttt{DynamicVariable<Z, T | NamedData>} & Represents a variable that is dynamically chosen, and cannot be used for implementing rules of algebra between variables.&&$\checkmark$ \\
\texttt{Term<T | NamedData | Variable, X>} & Defines the independent variable of the expression and points to the underlying data array (such a phase-field variable), and specifies a compile-time constant exponent \lstinline|X| applied to that variable. &&$\checkmark$\\
\hline
\end{tabular}
\end{table}

\newpage

\begin{table}[tb]
\small
\centering
\caption{Symbolic expression types implemented in \SymPhas{} (v1) and \SPnew{} (v2), and their roles in symbolic algebra. 
The prefix \texttt{Op} stands for operator.
\label{table:expression-evaluable}}
\begin{tabular}{|p{0.28\linewidth}|p{0.67\linewidth}|l|l|}
\hline
\textbf{Symbol / Node} & \textbf{Description} & v1 & v2 \\
\hline
\texttt{\small OpAdd<E...>} & \small Symbolic addition between any number of term \texttt{E}s. Addition is assumed to always be commutative. && $\checkmark$ \\
\texttt{\small OpBinaryAdd<L, R>}, \texttt{OpBinarySub<L, R>} & \small Symbolic addition between two operands of type \texttt{L} and \texttt{R}. Always be commutative. & $\checkmark$& \\
\texttt{\small OpBinaryMul<L, R>} & Symbolic multiplication of left expression of type \texttt{L} and right expression of type \texttt{R}. In general, this is a non-commutative operation to accommodate vector types. Commutativity is explicitly implemented as a rule for expressions which are known to be scalar-valued.& $\checkmark$& $\checkmark$ \\
\texttt{\small OpBinaryDiv<L, R>} & \small  Symbolic division of left expression of type \texttt{L} and right expression of type \texttt{R}. This is a non-commutative operation. & $\checkmark$& $\checkmark$\\
\texttt{\small OpTerms<Term...>} & \small A container of one or more objects of type \texttt{Term} (see Table~\ref{table:expression-terms}), representing the product between them. && $\checkmark$\\
\texttt{\small OpLVariable<G>}, \texttt{OpNLVariable<G...>} & \small A container representing either a single variable or product of variables, respectively. &$\checkmark$&\\
\texttt{\small OpIntegral<E>} & \small Symbolic representation of an integral of an expression of type \texttt{E}. &&$\checkmark$\\
\texttt{\small OpDerivative<D, E>} & \small General symbolic derivative, representing applying the differential operator of type \texttt{D} to the expression of type \texttt{E}. &&\\
\texttt{\small OpConvolution<L, R>} & \small Symbolic convolution operation between terms of type \texttt{L} and \texttt{R}. &$\checkmark$&$\checkmark$\\
\texttt{\small OpFunction<F, S...>}, \texttt{OpCallable<F, S...>} & \small User-defined or named function \texttt{F} taking arguments of type \texttt{S} that can be evaluated within the symbolic algebra expression. &&$\checkmark$\\
\texttt{\small OpSum<S>} & \small  Reduction operator that performs summation \texttt{S} over a list of variables. &&$\checkmark$\\
\texttt{\small OpChain<L, R, E>} & \small Symbolic chaining of operators (e.g., chained derivatives) of type \texttt{L} and \texttt{R}, applied to an expression of the type \texttt{E}. Chaining more than two operators is represented by nesting a chaining operation into either \texttt{L} or \texttt{R}. The types \texttt{L} and \texttt{R} may be non-evaluable, as long as applying them to \texttt{E} results in an evaluable expression. &&$\checkmark$\\
\texttt{\small OpCombination<L, R, E>} & \small Applying the addition of two operators of type \texttt{L} and \texttt{R} to the expression \texttt{E}. The types \texttt{L} and \texttt{R} may be non-evaluable, as long as applying them to \texttt{E} results in an evaluable expression. &&$\checkmark$\\ 
\texttt{\small OpPow<B, E>} & Exponentiation between expressions, where \texttt{B} is the base type, and \texttt{E} is the exponent type  (e.g., \texttt{B\^{}E}). &&$\checkmark$\\
\texttt{\small OpExponential<E>} &\small  A specialization of \texttt{OpPow}, where the base is the natural number $e$, and the expression of type \texttt{E} is the exponent. &$\checkmark$&$\checkmark$\\
\texttt{\small OpAbs}, \texttt{OpSqrt}, \texttt{OpSin}, \texttt{OpCos}, \texttt{OpTan} & \small Standard mathematical functions. &$\checkmark$&$\checkmark$\\
\hline
\end{tabular}
\label{table:expression-evaluable}
\end{table}

\begin{table}[tb]
\centering
\caption{Symbolic expression types implemented in \SymPhas{} and their roles in symbolic algebra. These types can be evaluated to yield a result. Most of constant types come from the previous version with the exception of \lstinline|OpTensor| and \lstinline|OpCoeff|. The columns v1 and v2 refer to the original version of \SymPhas{} and \SPnew{}, respectively.}
\label{table:expression-constants}
\small
\begin{tabular}{|p{0.3\linewidth}|p{0.5\linewidth}|l|l|}
\hline
\textbf{Symbol / Node} & \textbf{Description} & v1 & v2 \\
\hline
\texttt{\small OpLiteral<T>} & Represents numeric literal constants embedded in expressions, where \texttt{T} is the value type. &$\checkmark$&$\checkmark$\\
\texttt{\small OpIdentity} & Multiplicative identity, equal to 1. &$\checkmark$&$\checkmark$\\
\texttt{\small OpNegIdentity} & Negative of the multiplicative identity, equal to -1. &$\checkmark$&$\checkmark$\\
\texttt{\small OpVoid} & Additive identity, equal to 0. &$\checkmark$&$\checkmark$\\
\texttt{\small OpFractionLiteral<A, B>} & A compile-time fixed value of the fraction of natural numbers \texttt{A} and \texttt{B}. When the denominator is 1, it represents a  natural number. &&$\checkmark$\\
\texttt{\small OpNegFractionLiteral<A, B>} & The negative of a compile-time fixed value of a fraction. When the denominator is 1, it represents a negative integer. &&$\checkmark$\\
\texttt{\small OpTensor<T, Ns...>} & Represents an entry of type \texttt{T} (which can be any literal type) in the tensor defined by the numbers \texttt{N}. Currently, \SymPhas{} implements tensors for row and column vectors. &&$\checkmark$ \\
\texttt{\small OpCoeff<T>} & Contains an array of numeric literals of type \texttt{T} and represents the literal that is dynamically pointed to within the array. Used for models where the number of phase-fields is dynamically specified. &&$\checkmark$ \\
\hline
\end{tabular}
\end{table}

\begin{table}[tb]
\centering
\caption{Symbolic expression types implemented in \SymPhas{} and their roles in symbolic algebra. These types are not evaluable. All operators are new for the newest release. The columns v1 and v2 refer to the original version of \SymPhas{} and \SPnew{}, respectively.}
\label{table:expression-derivatives}
\small
\begin{tabular}{|p{0.3\linewidth}|p{0.5\linewidth}|l|l|}
\hline
\textbf{Symbol / Node} & \textbf{Description} & v1 & v2 \\
\hline
\texttt{\small OpOperatorDerivative} & Represents the operation of applying a derivative of order $p$. Mathematically, this is equivalent to $\nabla ^ p$ when $p$ is even, or $\nabla ^{p-1} \nabla$ when $p$ is odd. & &$\checkmark$\\
\texttt{\small OpOperatorDirectional\newline Derivative} & Represents the operation of applying a directional derivative of order $p$. Mathematically, this is $\partial^p /\partial u^p$, where $u$ is the axis $x$, $y$ or $z$. &&$\checkmark$\\
\texttt{\small OpOperatorMixed\newline Derivative} & Represents an operation of applying mixed derivatives, which are derivatives of different orders along different axes. A mixed derivative along a single axis is equivalent to a directional derivative. &&$\checkmark$\\
\texttt{\small OpOperatorChain<L, R>} & Represents chaining two operators, meaning that operator or expression of type \texttt{L} is applied to expression of type \texttt{R}. In general, \texttt{L} and \texttt{R} are not commutative. &&$\checkmark$ \\
\texttt{\small OpOperator\newline Combination<L, R>} & Represents adding two operators, meaning that operator or expression of type \texttt{L} is added to expression of type \texttt{R}. This operation should always commute. && $\checkmark$\\
\hline
\end{tabular}
\end{table}

\begin{table}[tb]
\centering
\caption{Non evaluable expression types implemented in \SymPhas{} and their roles in symbolic algebra. These types are used to implement symbolic algebra logic and construct evaluable expressions. The columns v1 and v2 refer to the original version of \SymPhas{} and \SPnew{}, respectively.}
\label{table:expression-symbols}
\small
\begin{tabular}{|p{0.3\linewidth}|p{0.5\linewidth}|l|l|}
\hline
\textbf{Symbol / Node} & \textbf{Description} &v1&v2 \\
\hline
\texttt{\small Symbol} & Named symbolic variable, typically used in functions or summations.&&$\checkmark$ \\
\texttt{\small SymbolicDerivative} & Partial derivative with respect to a symbol. The symbolic algebra library can symbolically apply these derivatives at compile time to arbitrary expressions. This is distinct from the derivative expressions listed in Tables~\ref{table:expression-evaluable} and~\ref{table:expression-derivatives}, which are derivatives of phase-field variables, and represent numerical derivatives, for example via stencils. &&$\checkmark$\\
\texttt{\small SymbolicFunctional\newline Derivative} & Variational (functional) derivative; used to derive equations of motion from free energy.&&$\checkmark$ \\
\texttt{\small SymbolicSum} & Symbolic summation over indexed expressions. &&$\checkmark$\\
\texttt{\small DynamicIndex} & Symbolic index in a list of variables of dynamic length. &&$\checkmark$\\
\hline
\end{tabular}
\end{table}

\end{document}